\documentclass[11pt]{aastex}
\usepackage{graphicx, emulateapj5, apjfonts}
\usepackage{onecolfloat}
\usepackage{epsf}

\shortauthors{Aday R. Robaina et al.}
\shorttitle{The merger-driven evolution of massive galaxies}

\begin{document}

%%%%% Added the \def\head{ lark.

\def\head{

\title{The merger--driven evolution of massive galaxies}

\author{Aday R. Robaina$^1$,
Eric F.\ Bell$^2$,
Arjen van der Wel$^1$,
Rachel S. Somerville$^3$,
Rosalind E.\ Skelton$^1$,
Daniel H. McIntosh$^4$,
Klaus Meisenheimer$^1$ \&
Christian Wolf$^5$\\}

\affil{$^1$ Max-Planck-Institut f\"ur Astronomie, K\"onigstuhl 17, D-69117 Heidelberg, Germany; \texttt{arobaina@mpia.de}\\
$^2$ Department of Astronomy, University of Michigan, 500 Church Street, Ann Arbor, MI 48105 USA \\
$^3$ Space Telescope Science Institute, 3700 San Martin Dr., Baltimore, MD
21218, USA\\
$^4$ Department of Physics, University of Missouri-Kansas City, Kansas City,
  MO 64110,  USA \\
$^5 $ Department of Physics, Denys Wilkinson Bldg., University of Oxford, Keble Road, Oxford, OX1 3RH, UK
}

\begin{abstract} We explore the rate and impact of galaxy mergers on the massive galaxy population using the amplitude of the two-point correlation function on small scales for $M_*>5\times 10^{10}M_\sun$ galaxies from the COSMOS and COMBO--17 surveys. Using a pair fraction derived from the {\it Sloan
  Digital Sky Survey} as a low-redshift benchmark, the large survey area at
intermediate redshifts allows us to determine the evolution of the close pair
fraction with unprecedented accuracy for a mass-selected sample: we find that
the fraction of galaxies more massive than $5\times 10^{10} M_\sun$ in pairs separated by less than 30~kpc
  in 3D space evolves as $F(z)=(0.0130\pm0.0019)\times (1+z)^{1.21\pm0.25}$
  between $z=0$ and $z=1.2$. Assuming a merger time scale of 0.5
  Gyrs, the inferred merger rate is such that galaxies with mass in excess of
  $10^{11}M_\sun$ have undergone, on average,  0.5 (0.7) mergers involving
  progenitor galaxies both more massive than $5\times 10^{10}M_\sun$ since
  $z=$~0.6 (1.2). We also study the number density evolution of massive 
  red sequence galaxies
  using published luminosity functions and constraints on the $M/L_B$
  evolution from the fundamental plane. Moreover, we demonstrate that the measured 
  merger rate of massive galaxies is sufficient to explain this observed number density evolution in
  massive red sequence galaxies since $z=1$.

\end{abstract}

\keywords{galaxies: general --- galaxies: evolution --- galaxies: interactions
  --- galaxies: ellipticals and lenticulars, cD --- galaxies: formation --- galaxies: statistics
}
}%%%end head

\twocolumn[\head]

\section{Introduction}

The distribution of galaxies in a colour-magnitude diagram shows a bimodality:
galaxies with red optical colors occupy a relatively tight sequence in color (the
red sequence) while blue galaxies show a wider dispersion and populate the
so-called `blue cloud' \citep{blanton03}. This bimodality has existed for most
of the history of the universe \citep[e.g.,][]{bell04, faber07,
  taylor}. However, the stellar mass density of red sequence galaxies has
increased by a factor $\sim 2$ since $z=1$, while it has remained
approximately constant for blue galaxies \citep{bell04,faber07,brown}. Yet,
the majority of stars are formed in blue galaxies \citep{bell07, walcher},
which implies that galaxies migrate from the blue cloud to the red sequence
\citep[see, e.g., the discussion by ][]{faber07}.

While significant evolution of the integrated mass function of red galaxies is
generally agreed upon, the evolution in the number density of high-mass
galaxies has been shown to be consistent with zero \citep[e.g.,][]{cimatti,
  cool}. However, given the large uncertainties, the constraints are not
tight. There is ample room for considerable evolution, and such evolution is
expected in a $\Lambda$CDM hierarchichal cosmology, which predicts significant
growth of galaxies through merging, especially at the high-mass end
\citep{delucia07}.

The most massive, non-starforming galaxies are essentially all spheroidal
(i.e., have ellipticities < 0.4; \citealp{arjen}), indicating that they formed through merging \citep{toomre, sch,
  kormendy09}. Shape distributions do not specify {\it when}
merging happened.  Some valuable constraints have been placed by the study
of the clustering properties of red galaxies in this mass range by
\citet{white}, who found that 1/3 of massive red satellites disappear between
z=0.9 and z=0.5 by merging with the central galaxies in their halos.

The goal of this paper is to estimate the impact of galaxy mergers on the
 evolution of massive galaxies from $z\sim1$ to the present day, and more
   specifically, on the massive end of the red sequence. To that end we
 measure the merger fraction since $z=1.2$ in a large sample of mass-selected
 galaxies. We will test whether these measurements are consistent with
 predictions from hierarchichal formation models, and, providing an
 independent estimate of the number density of the most massive red galaxies,
 with the observed evolution of the bright tail of the luminosity function of
 red galaxies.

It is observationally challenging to identify galaxy mergers, especially at
large cosmological distances. Mergers are found
either in an early phase of the interaction, when the two galaxies have not
yet coalesced and are found in a close pair \citep{patton00, lefevre, lin04, lin08, bell06a, karta, robaina09}, or in
a later phase, when they display signatures of gravitational
interaction and are just prior to or after coalescence \citep{abraham, cas, lotz,
  bell05, mcintosh, jogee, amanda, robaina09}. A wide range of results have been found for the merger
fraction evolution, parameterized as
$(1+z)^m$, with $m$ ranging from 0 to 4 \citep[e.g.,][]{patton00,lefevre, lin04,karta}.

Here, we
  use robust 2--point correlation function techniques on a sample of galaxies with $0.2<z<1.2$ with
stellar masses in excess of $5\times 10^{10} M_\sun$ selected from the COSMOS and
COMBO--17 surveys to quantify the merger rate of massive galaxies and its
evolution. We augment the statistical significance of our analysis by using an estimate of the pair fraction found in Sloan Digital
Sky Survey (SDSS) at $z\sim 0.1$. The total volume probed by this study at intermediate
redshifts is at least 4 times larger than any previous mass--selected study on the evolution
of the merger fraction, and reduces dramatically the systematic uncertainties
related to cosmic variance by the use of four independent fields. Then we compare the inferred galaxy merger rate with the
observed number density evolution of massive ($M_*>10^{11}M_{\sun}$), red
galaxies from $z\sim 1$ to the present day, which we obtain by converting
\citet{brown} LFs to stellar--mass functions.
We assume $H_0=70$~km/s, $\Omega_{m}=0.3$ and $\Omega_{\Lambda}=0.7$.

\section{Sample and method}\label{sec:sample}

The bulk of our sample is drawn from the $\sim$2\,sq.\ degree 
COSMOS survey \citep{scoville}.  We use photometric redshifts calculated
from 30-band photometry by \citet{ilbert}; comparison with 
spectroscopic redshifts shows excellent accuracy. 
We use those redshifts to derive
rest-frame quantities and stellar masses by using the observed broad-band
photometry in
conjunction with a non-evolving template library
 derived using P\'egase stellar population model \citep[see][]{fioc99} and a
 \cite{chabrier03} stellar initial mass function (IMF). The use of a  \cite{kroupa93}
 or a \cite{kroupa01} IMF would yield similar stellar masses to within $\sim$
 10\%. The reddest templates are produced through single
 exponentially--declining star formation episodes, intermediate templates also
 have a low-level constant star formation rate (SFR) and the bluer templates
 have superimposed a recent burst of star formation. A full description of the
 stellar masses will be provided in a future paper; 
comparison with \citet{pan} masses shows agreement at the 0.1\,dex level.

To combat the sample variance of a single 2 sq.\ degree field, 
we augment the COSMOS sample with a sample drawn from three 
widely-separated 0.25 sq.\ degree fields from COMBO-17. 
COMBO--17 photo-z's, colors and stellar masses have been extensively
described in \citet{wolf03},\citet{wolf04}, \citet{borch} and \citet{gray}. Given the different
depths of the two surveys we only include galaxies from the COMBO-17 catalog
at $z<0.8$, where it is complete for our mass limit.
The final sample comprises $\sim 18000$ galaxies with $M_*\ge 5\times
10^{10}M_\sun$ over an area of $\sim 2.75$ sq. deg. in the redshift range $0.2<z<1.2$.

We use the fraction of galaxies with a
companion closer than 30 kpc (close pairs) as a  proxy for the merger fraction, as those systems are very likely to merge in a
few hundred Myr\footnote{As the lower mass limit of our sample is $5\times
  10^{10}M_\sun$, we are automatically selecting both members of the pair to be
  above that mass.}. As redshift errors translate into line-of-sight (l.o.s.) distance
uncertainties of the order of $\sim 50-100$~Mpc we use projected 2-point
correlation functions (2pcf) to find the number of projected close pairs and then deproject into the 3D space.

The projected correlation function $w(r_P)$ is the integral along the line of
sight of the real-space correlation function:

\begin{equation}
w(r_{p})=\int^{\infty}_{-\infty} \xi([r_p^2+\pi^2]^{1/2})d\pi,
\end{equation}
where $r_{p}$ is the distance between the two galaxies projected on the plane
of sky and $\pi$ the line-of-sight separation. A convenient and simple
estimator of the 2pcf at small scales is $w(r_{p})=\Delta(DD/RR-1)$ \citep[e.g.][]{bell06a,robaina09}, where $\Delta$
is the path length being integrated over, $DD(r_P)$ is the histogram of separations
between real galaxies and $RR(r_P)$ is the histogram of separations between galaxies
in a randomly-distributed catalog. As shown in the literature
\citep[e.g.][]{davis, li}, the real-space 2pcf can be reasonably well fit by a
power-law. Assuming $\xi(r)=(r/r_0)^{- \gamma}$, then
$w(r_p)=Cr_{0}^{\gamma}r_{0}^{1-\gamma}$, with
$C\sqrt{\pi}\{\Gamma[(\gamma-1)/2]/\Gamma(\gamma/2)\}$. We fit the latter
expression to our data to find the parameters $\gamma$ and $r_0$ and use them
in the real-space 2pcf to find the fraction of galaxies in close pairs.

As we wish to preserve S/N we did not integrate along
the entire l.o.s. when calculating $w(r_P)$. Instead we allowed
galaxies to form a pair only if the redshift difference was smaller than
$\sigma_{pair}=\sqrt{2}\times \sigma_z$, with $\sigma_z$ being the redshift
error of the primary galaxy. As the photo-z errors follow a Gaussian
distribution with width $\sigma_z$ \citep{wolf03,wolf04} a fraction of the
pairs would be missing by simply imposing the l.o.s. criteria. Thus, following
\citet{bell06a} the
fraction of galaxy pairs is

\begin{equation}
f=\int^{\sigma_{pair}}_{- \sigma_{pair}}\frac{1}{\sqrt{2\pi \sigma_{z}}}e^{-z^2/2\sigma_{z}^2}dz.
\end{equation}

Then, $w(r_P)$ is multiplied by $1/f$ in order to account for missing pairs; in our case
a correction factor of 1.19.

Given the 3D correlation function $\xi(r)$, the differential probability of finding a galaxy occupying a volume $\delta V$
at a distance $r$ of another galaxy is $\delta P=n[1+\xi(r)]\delta V$, where $n$ is the number density of secondary
galaxies. Then, by a simple integration of this expression, we obtain
the probability of a galaxy being within a distance $r_f$ of any other galaxy
\citep{patton00, bell06a, masjedi06}:

\begin{equation}
P(r\le r_f)=\int^{r_f}_{0}{n[1+\xi(r)]dV}
\end{equation}

Because $\xi(r)=(r/r_0)^{-\gamma}$ and $\xi(r)\gg 1$ at the small
scales, we obtain:

\begin{equation}
P(r\le r_f)=f_{pair}=\frac{4\pi n}{3-\gamma}r_0^\gamma r_f^{3-\gamma},
\end{equation}
where $f_{pair}$ is the fraction of {\it galaxies} in close pairs.
As galaxy interactions are completely decoupled from the Hubble flow, we
calculate probabilities as a function of the proper (physical) separation
between the two galaxies.
Errors in the correlation function are calculated by means of bootstrap resampling.

\section{Results and discussion}\label{sec:results}

\begin{table*}[ht!]
\begin{center}

\caption{3D correlation function parameters and close pair fractions {\label{tab:cor}}}
\small{
\begin{tabular}{cccccc}

\hline
$z$ & $\gamma$ & $r_0$ & $n(Mpc^{-3})$& $f_{pair}$ (< 30~kpc) \\
\hline
$0.2<z\le0.4^a$ & $2.03\pm 0.05$ & $3.50\pm 0.50$ & 0.0031 & $0.0171\pm 0.0050$\\
$0.4<z\le0.6^a$ & $2.06\pm 0.04$ & $3.60\pm 0.50$ & 0.0034 & $0.0238\pm 0.0043$\\ 
$0.6<z\le0.8^a$ & $1.94\pm 0.04$ & $3.80\pm 0.30$ & 0.0064 & $0.0241\pm 0.0038$\\
$0.8<z\le1.0^b$ & $1.93\pm 0.03$ & $2.85\pm 0.25$ & 0.0131 & $0.0277\pm 0.0031$\\
$1.0<z\le1.2^b$ & $1.96\pm 0.03$ & $2.55\pm 0.20$ & 0.0162 & $0.0315\pm 0.0030$\\
\hline\\

\multicolumn{4}{l}{$^a$ Combined COSMOS+COMBO--17 sample}\\
\multicolumn{4}{l}{$^b$ COSMOS--only sample}\\

\end{tabular}}
\end{center}

\end{table*}

\begin{figure}[h!]

\begin{center}
\includegraphics[width=8.5cm,height=7.5cm]{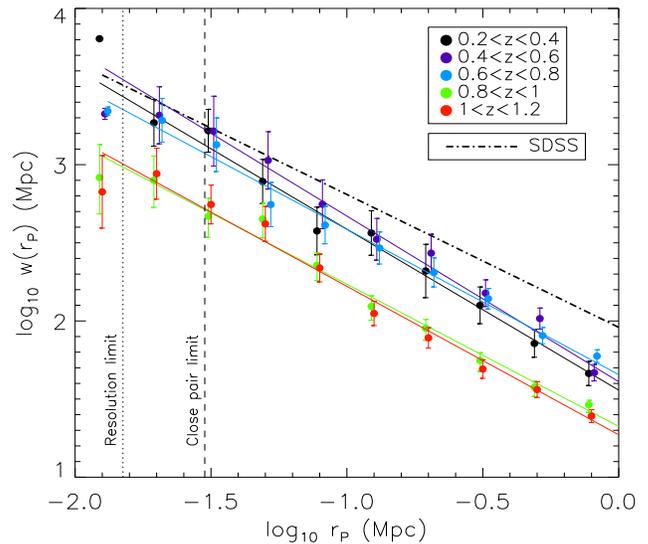}
\caption{Projected 2--point correlation functions for galaxies with masses
  $ \ge 5\times 10^{10} M_\sun$ from our combined COSMOS + COMBO--17 sample in
  different redshift bins. The best--fit power law is overplotted following
  the color code of the different redshift bins. The power-law parameters are
  given in Table~\ref{tab:cor}. The dot-dashed line represents the
  best--fit power law to the SDSS sample derived from results by
  \citet{li}. The vertical dashed line represents the separation limit for
  close pairs of 30~kpc (physical). The vertical dotted line represents the
  (average of the redshift bins) limit where one can no longer reliably resolve
  individual galaxies in a pair from the ground--based photometry. 
}\label{fig:cors}
\end{center}
\end{figure}

This work aims to an understanding of the impact of galaxy mergers on the
evolution of massive galaxies and the evolution of the red sequence at its
massive end. To that end we apply the above method to our sample of galaxies
with $M_*>5\times10^{10}M_\sun$, regardless of their optical colors, to find
the fraction of galaxies in close physical pairs, which we will use as a proxy
for the fraction of galaxies undergoing a galaxy interaction. Hereafter, all reference to
close pairs will refer to galaxy pairs with inferred separations smaller than 30 kpc in
3D space. Fig.~\ref{fig:cors} and Table~\ref{tab:cor} show the result of our
2pcf analysis.

 The projected
correlation functions in Fig.~\ref{fig:cors} are shown over the range $10<r_p/kpc<1000$,
in physical (not comoving) units, as is appropriate for a process such as
galaxy merging that has decoupled from the Hubble flow. One can see that the
fitted power laws are good descriptions of the run of $w(r_p)$ with radius,
down to the resolution limit of  $\sim 15$\,kpc. Below this radius, we
extrapolate the power law fit. We have checked the accuracy of this
extrapolation by comparison with extremely close pairs resolved by HST in the
study of two of the four fields used in this paper by \citet{robaina09}. An
extrapolation of our power-law fits to small radius would predict that $\sim
40\%$ of all our galaxies in $r<30$~kpc pairs should have $r<15$~kpc separations; we find
that $\sim30\%$ of galaxies in close pairs have $r<15$~kpc separations using
HST data. Accounting for the fact that \citet{robaina09} required additional
interaction signatures  such as tidal tails for a pair to be assigned to that category\footnote{While most of the projected close pairs at $r<15$~kpc
    are likely to be undergoing an interaction, the morphological signatures
    do not need to be necessarily evident. Gas fractions and orbital
    parameters strongly influence the strength of the tidal features. Then, we
    would expect to find a fraction of extremely close pairs (respect to the
    total number of close pairs with $r_p<30$~kpc) slightly below the expected
    40\%.}, and that some extremely close pairs will be classified as already
  coalesced merger remnants, we argue that this level of agreement is
  remarkable, and we conclude our extrapolation to radii $r<15$~kpc is appropriate\footnote{Note that we
could have seemingly sidestepped this issue by using pairs with
$15<r<30$\,kpc for the close pair measurement, and adjusting the timescale
to reflect the amount of time pairs spend separated by $15<r/kpc<30$. 
Nonetheless, in this case, one would still want to confirm that systems
that reached 15\,kpc continue to approach smaller radii, amounting to an exercise
equivalent to confirming the power law extrapolation.}.

In Table~\ref{tab:cor} we show the parameters for the 3D correlation function
that we obtain by deprojecting the 2D 2pcf. The evolution of the merger fraction is
mainly driven by the evolution in the number density of galaxies above
$5\times 10^{10}M_\sun$ in physical coordinates, together with some evolution
in the clustering length $r_0$.

\begin{figure}[h!]

\begin{center}
\includegraphics[width=8.5cm,height=7.5cm]{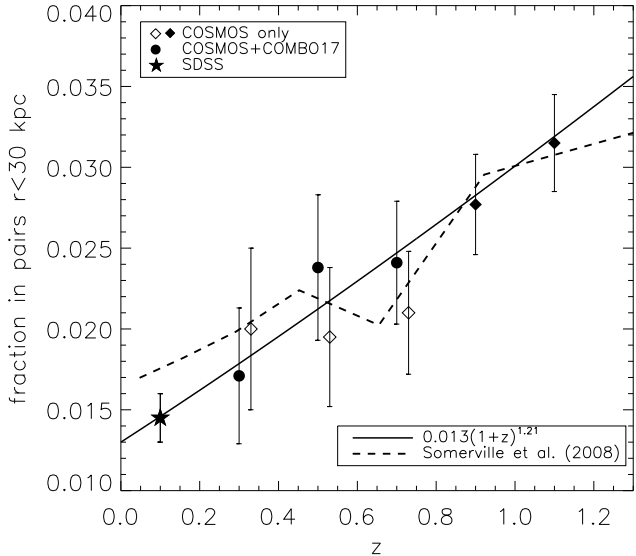}
\caption{Fraction of $M_*>5\times 10^{10}M_\sun$ galaxies in
  close (3D) pairs ($r<30$~kpc)
as a function of redshift. {\it Diamonds:} Pair fraction found using only the
COSMOS catalog. {\it Black circles:} Pair fraction found when adding galaxies from
the COMBO--17 survey at $z<0.8$ (we do not include galaxies from the COMBO--17
catalog in the two higher $z$ bins). {\it Star:} Pair fraction from
SDSS. The line shows the best fit to a real space pair fraction evolution with shape $F(z)= f(0)\times(1+z)^m$, with
 $f(0)=0.0130\pm 0.0019$ and $m=1.21\pm 0.25$ (fit to all black-filled points:
 star, circles and diamonds). {\it Dashed line:} Fraction of
 galaxies above $M_*>5\times 10^{10}M_\sun$ involved in mergers, predicted by
 \citet{somer} models when assuming a timescale of 0.5 Gyrs to transform from
 merger rate to fraction.
}\label{fig:fraction}
\end{center}
\end{figure}

In Fig.~\ref{fig:fraction} 
we show the fraction of massive galaxies found in pairs
with separations $r<30$~kpc as a function of $z$. To augment the data
at $z \sim 0.1$, we calculate the SDSS close pair fraction using 
 $\gamma$ and $r_0$ given by \citet{li}. We adjust their $r_0$ by $\sim 5$\% (an empirical adjustment based on the COSMOS+COMBO-17 sample) to account for different lower mass limits and binning.  We also use  the number density of galaxies fulfilling our
mass criteria, adopting the g-band selected stellar-mass function in
\citet{bell03} after correcting for stellar IMF and $H_0$.

We perform an error-weighted least-squares fit of the form $F(z)= f(0)\times (1+z)^m$ to the filled
points in Fig.~\ref{fig:fraction}, i.e., the SDSS pair fraction, the combined
COSMOS+COMBO--17 pair fraction at $0.2<z<0.8$ and the COSMOS pair fraction at
$0.8<z<1.2$. We find $f(0)=0.0130\pm 0.0019$ and $m=1.21\pm 0.25$. These associated uncertainties are at the
same level as the systematic uncertainties in close pair fraction evolution
caused by uncertainty in the overall stellar M/L scale, and its evolution
since $z=1$.

In our correlation function we
do not impose a specific mass ratio criteria, but given the shape of the mass
function above $M_*>5\times10^{10}M_\sun$, we expect most of the mergers to be
majors; i.e., with mass rations between 1:1 and 1:4. We measure the fraction
of pairs in projection which fulfill such a mass ratio criterion, finding
 a value between 70 and 90 percent in the different redshift bins.

Galaxies from the COSMOS survey represent 70\% of our sample in the bins
ranging from $z=0.2$ to $z=0.8$, however, the addition of COMBO--17
galaxies helps to decrease the sample variance. From the work by 
\citet{moster}, we estimate that the sample variance is reduced by a factor of
$\sim 30\%$ by including the three $\sim 0.25$~sq. deg independent fields from
COMBO--17. This effect is clearly seen in Fig.~\ref{fig:fraction}. Considering
only galaxies from COSMOS, there is an abrupt transition between $z=0.7$ and
$z=0.9$, which is smoothed by the inclusion of COMBO--17 galaxies.

%rss rewrote this pp
In Fig.~\ref{fig:fraction} we also show the prediction for the
close-pair fraction from the semi-analytic models (SAM) of \citet{somer},
updated as described in \citet{hopkins09}. The models shown here
adopt the best-fit cosmological parameters given in
\citet{komatsu}. We extract the merger rate of galaxies where both
progenitors have stellar masses above $5\times10^{10}M_\sun$, normalize by the
total number density of galaxies above this mass and multiply by two to obtain
the fraction of galaxies involved in mergers (the quantity that we estimate
observationally), per Gyr. Then we multiply by the timescale over which a merger would be observed
as a ``close pair'', which here we assume to be $\tau = 0.5$
Gyr following \citet{patton08} and \citet{bell06a} in order to convert this merger rate into the equivalent of the observed fraction. With this assumed
timescale the model predictions are in
excellent agreement with our observational results. Assuming an uncertainty of
$\sim 30\%$ in the timescale (matching the numbers from \citealp{lotz09} for close galaxy
  pairs timescales) would correspondingly shift vertically the model line by a
  similar number. Predictions from
other SAMs tend to agree reasonably well with those of the
\citet{somer} models, as shown in Fig.\ 11 of Jogee et al. (2009; see
also Guo \& White 2008 and \citealp{hopkins10}).

\subsection{Comparison with previous works}

\begin{figure}[h!]

\begin{center}
\includegraphics[width=8.5cm,height=7.5cm]{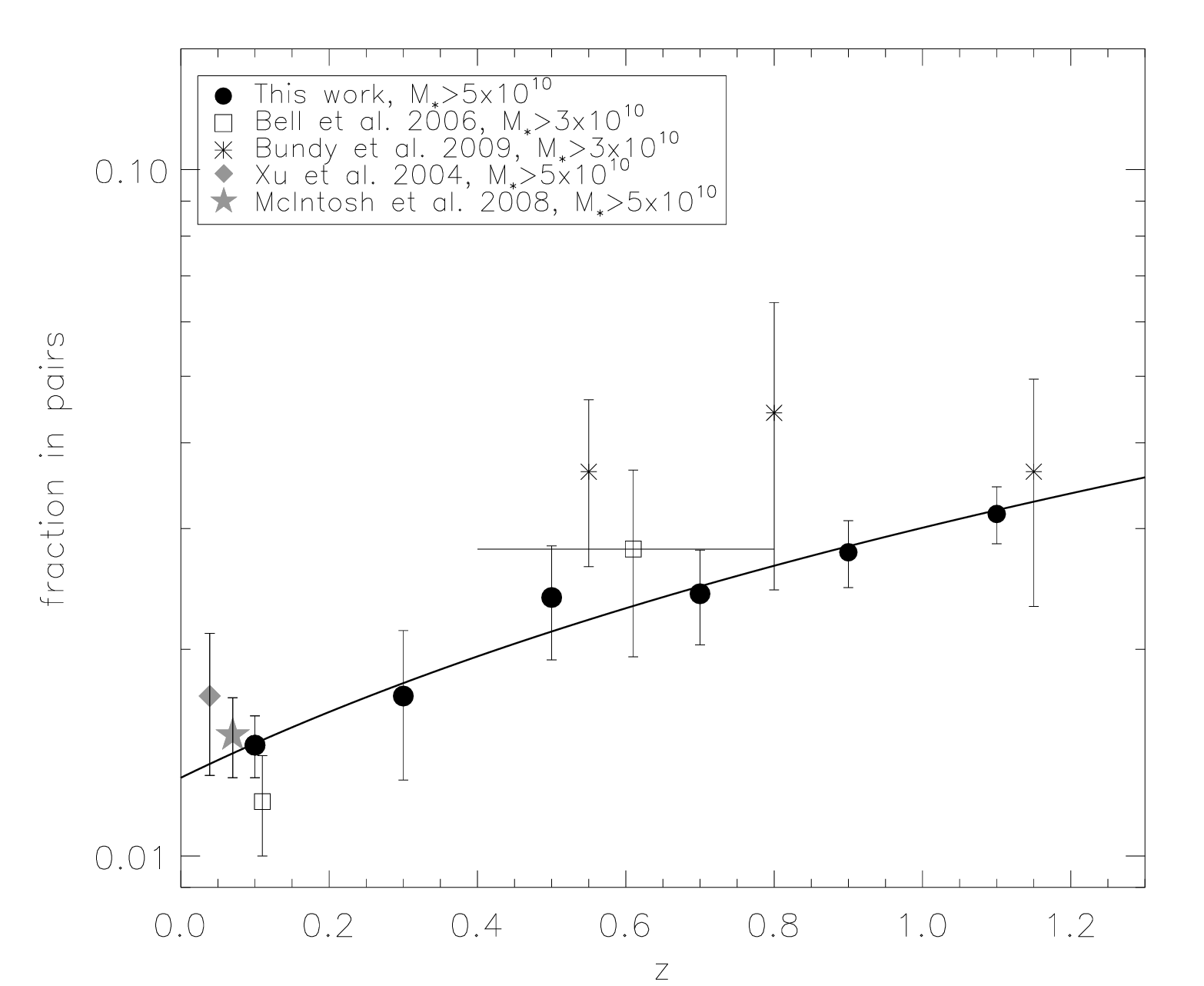}
\caption{\label{fig:comp} Evolution of the number of galaxies in close pairs. The estimate for major pair fraction of
  galaxies $M>3\times10^{10}M_{\sun}$ from \citet{bell06a} is shown as empty
  squares. The point
  by \citet{xu} is shown as the grey diamond. \citet{mcintosh} is shown as the
  grey star. The results of \citet{bundy} have been corrected down by 33\% in
  order to match our pair definition (see text for details).
}
\end{center}
\end{figure}

Very few studies of close pair fraction evolution use 
mass-limited samples. \citet{bell06a} used COMBO--17 data (0.75 sq.~deg., the
same catalog we use here to complement our COSMOS catalog) to find a fraction
of galaxies in close major pairs ($r<30$~kpc) of 2.8\% for galaxies more
massive than $3\times 10^{10}M_{\sun}$ at $0.4<z<0.8$, in excellent agreement
with our result despite the slightly different mass limit. They also
performed an autocorrelation of all galaxies with $M_*>2.5\times
10^{10}M_\sun$ finding a pair fraction of $\sim 5\%$. Adopting in our analysis 
instead a mass limit of $M_*>2.5\times
10^{10}M_\sun$ we recover a pair fraction of $\sim 5$\%. The 
main driver of this higher pair fraction is the fact that $\sim 50$\% of 
close pairs in a sample limited to have  $M_*>2.5\times
10^{10}M_\sun$ have mass ratios between 1:4 and 1:10; i.e., the major 
merger fraction is similar, but the close pair fraction is boosted by 
a considerable contribution from minors. 

Recently, \citet{bundy} performed a similar measurement by
studying a sample of galaxy pairs from the GOODS fields (total area $\sim 0.1$
sq. deg.). In Fig.~\ref{fig:comp} we show their results for the mass range $>3\times 10^{10}M_\sun$
after converting their fraction of pairs to the fraction of {\it galaxies in
  pairs}. \citet{bundy} study the fraction of major mergers in 
galaxies with $M_* > 3\times 10^{10}M_\sun$, including galaxies 1.5 mags
fainter than that mass limit, while by our definition of a close pair, both
galaxies need to be above our mass limit. It is natural that they find a
higher pair fraction, but we quantify this difference by performing a test in
which we try to reproduce the \citet{bundy} methodology. Given the nature of our method it is difficult
to perform a cross-correlation in order to match their pair definition;
instead we
study the 2D distribution of galaxies in a similar fashion as
\citet{bundy}. That is, we look for galaxies with separations $r<30$/kpc and
matched in redshift space and subtract random counts. We perform such a
test in the redshift range $0.2<z<0.6$, where we are complete for stellar
masses of $\sim 1-2 \times 10^{10}M_\sun$, with both our pair definition (both galaxies
above $5\times 10^{10}M_\sun$) and also keeping such a mass limit for the
primary but allowing the secondary to be a factor 4 less massive. In the first
case we find, as expected, a result compatible with the one found with the 2pcf,
and in the second we find that the pair fraction is larger by a factor of
$\sim 1.5$. Extrapolating this ratio to the higher redshift bins, we
accordingly re-scale the \citet{bundy} numbers by multiplying their results by 0.67 in order to match our
pair definition. The
difference is reduced to a factor of less
than 2 in the lower redshift bins, while our results basically agree at $z\sim
1$. We also show the estimate from \citet{xu}, who used a combined sample from the
2MASS and 2dFGRS surveys, after converting their results to our IMF and $H_0$,
and the result from \citet{mcintosh}, who measured the merger fraction in
galaxy groups based in projected pairs with additional signs of gravitational interaction.
Before performing the comparison, we further correct the pair fractions found by \citet{xu} and \citet{bundy}
down by 30\% to account for pairs in their analyses that are genuinely associated with each other (so are not random projections), have projected
separations of $<30\,$kpc but are separated by more than 30kpc in real space
(i.e., galaxies in groups that are projected along the line of sight; \citealp{bell06a})\footnote{This
  effect is present also in pair fractions determined from spectroscopic
  redshifts as long as a deprojection to the 3D space is not performed.}.

We can not compare our measurements with morphological studies of merger
fractions because of uncertainties with the nature of the progenitors (
  differentiation between major and minor mergers can be extremely
challenging, especially in the case in which one of
  the progenitors is very gas-rich), and
selection effects related to orbital parameters, galaxy structure and gas
fractions, with the latter introducing a bias in the recovery of interacting
systems at higher $z$ \citep{lotz09}. A more fundamental difference
  with respect to morphological studies, both
  those performed with visual classifications \citep{jogee} and with automated
  classifiers \citep{conselice09, lotz08}, is the fact that both merger
  remnants and closely interacting pairs can show signatures of gravitational
  interactions \citep[see][for a discussion on the relative numbers of merger
  remnants and close pair interactions with morphological
  disturbances]{robaina09}. These two groups are related to different time
  scales, and the strength and abundance of tidal features is also different (often
  degenerate with other parameters such as gas fraction). Studies on close
  pair statistics aim to detect the galaxy interaction in an early
  stage, where both progenitors can be studied separately, while morphological
studies include systems both before and after the coalescence. For this
reason, a direct, quantitative comparison is impossible. It is also hazardous to compare with close pair measurements based
on luminosity-selected samples. \citet{lin08} used a sample
of galaxies with B-band magnitudes (corrected for passive luminosity
evolution) $-21<M_B<-19$. For red galaxies, this is roughly compatible with
our mass--selected sample, but it includes many low-mass blue galaxies, which
makes a comparison impossible because their clustering properties are very
different. As an extreme example, in which one close pair is formed by a
  very red $M_B=-21$ galaxy and a very blue one with $M_B=-19$, the
  stellar-mass ratio in this hypothetical interaction could be as high as
  40. This pair definition is clearly not compatible with one in which both
  galaxies are selected by stellar mass. Furthermore, as merging can enhance the star formation
activity \citep{barton2000, robaina09}, selecting galaxies in rest-frame blue bands is
  biased in favor of mergers, as such a selection would recover merging systems
  with lower mass given the decreased M/L ratio induced by the
  interaction. All of these effects will be more pronounced in 
gas--rich galaxies, and gas fractions were likely higher in the past, 
leading to a redshift dependent bias.

\citet{karta} derived the evolution of the pair
fraction from a luminosity-selected sample from a similar dataset
as analyzed in this paper.  Their very different result (they
find $m=3.1\pm0.1$) is caused at least in part by the difference between
mass-- and luminosity--selected samples, as described above, as well as by the
fact that they do not correct for passive luminosity evolution.  In
addition, they identify pairs in both the ground-based and
HST/ACS-based catalogues. Because their sample is selected by
ground-based luminosity, very close pairs that are only resolved
by ACS can be as bright as a single galaxy in a more widely
separated pair that is resolved in the ground-based imaging.
This artificially raises that close pair fraction, especially at high
redshift.

Summarizing, we choose not to compare with merger fractions obtained by
  morphological indicators, nor with those obtained by close pair counts
  in luminosity-selected samples; these provide measurements of different
  merger-related quantities and can not, in general, be compared to our measurement.

\subsection{The impact of galaxy merging on the creation of
  $M_*>10^{11}M_\sun$ galaxies.}\label{sec:rate}

Our measured fraction of $M_*>5\times10^{10}M_\sun$ galaxies in close pairs as
a function of redshift constrains the impact that merging-induced mass assembly has on the creation rate of $M_*>10^{11}M_\sun$ systems.

As only $\sim6\%$ of our galaxies with a projected companion at $r<30$~kpc has a {\it second} companion at such separation, we
assume that the number of close pairs is simply one half of
the number of galaxies in close pairs, neglecting the very minor influence of
such a small number of galaxies in triplets. The fraction of newly created $M_*>10^{11}M_\sun$ galaxies due to merging is
$f_{rem}=N_{cp}/N_{11}$, where $N_{cp}$ is the number of
close pairs of galaxies above $5\times 10^{10}M_\sun$ each (regardless of
their optical colors) and $N_{11}$ is
the total number of galaxies with stellar masses in excess of $10^{11}M_\sun$.

Following \citet{patton08} or \citet{bell06a}, the merger timescale for galaxy pairs at this
separation is approximately $\tau=0.5$~Gyrs, so we compute the creation rate of newly
assembled galaxies, $R_{rem}$, as $R_{rem}=f_{rem}/\tau$. We integrate the
merger rate over cosmic time finding that, on average, present day galaxies with stellar masses
larger than $10^{11}M_\sun$ have undergone 0.5 (0.7) mergers since $z=$~0.6
(1.2) from interactions between galaxies more massive than $5\times
10^{10}M_\sun$. 

\subsection{The merger--driven evolution of the red sequence.}
In the previous sections we have studied the impact of merging of the
evolution of the mass function of galaxies, without restriction to the optical
colors of the progenitors or descendants. Now we want to address the question whether our observed merger rate evolution
can explain the observed number density evolution in the massive end
($M_*>10^{11}M_\sun$) of the red sequence.

\begin{figure}[h!]

\begin{center}
\includegraphics[width=8cm,height=7cm]{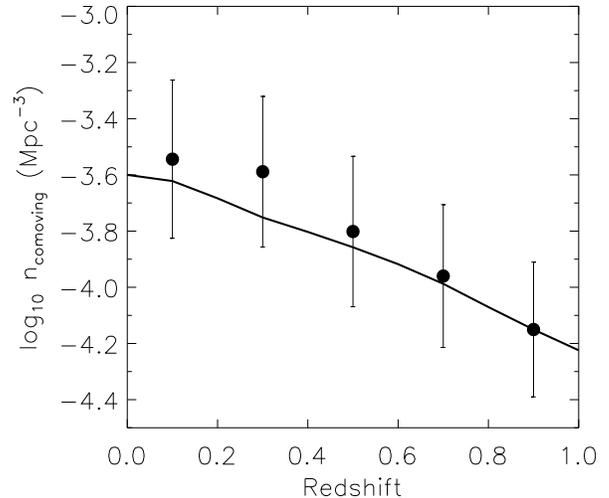}
\caption{\label{fig:evol} Number density evolution of red galaxies with
  $M_*>10^{11}M_\sun$. Filled points with error bars represent the
  number density observed by
  \citet{brown} when translated to stellar mass (see text for details). The solid line shows the expected growth
  implied by our measurement of the merger rate.
}
\end{center}
\end{figure}

First, we convert the evolution of the B-band luminosity function (LF) of red
galaxies as measured by Brown et al. (2007) to the evolution in the number
density of galaxies more massive than $10^{11}M_\sun$. For this conversion we use the typical stellar mass-to-light ratio of nearby red galaxies
($M/L_B=3.4\pm0.6$, e.g., \citealp{bell03, kauffmann}) and
apply a correction to account for its evolution with redshift, derived from
the evolution of the fundamental plane (FP) zero point ($\Delta
\log(M/L_B)=0.555 \pm 0.042$ between $z=1$ and $z=0$; \citealp{vd}). We repeat the process 10000
times allowing the luminosity
function parameters, as well as the $M/L_B$ constraints, to vary randomly
within their uncertainties. We estimate the final uncertainty by
estimating the typical dispersion of those 10000 realizations. The results
from this exercise are shown in
Fig.~\ref{fig:evol}. We find an
  evolution in the number density of the massive end of the red sequence of a
  factor $\sim 4$ between $z=1$ and $z=0$.

Our inference of a factor $\sim 4$ evolution in the number density of red
sequence galaxies with $M_*>10^{11}M_\sun$ is somewhat larger than a number of
recent estimates. The main reason for this is our adoption of the recent
results by \citet{vd} for the evolution of M/L determined through FP
evolution, who find relatively rapid evolution of M/L leading to more rapid
inferred fading of the stellar population than assumed by many previous
works. For example, \citet{cimatti} found a result compatible with a constant number density of $M_*>10^{11}M_\sun$
galaxies in the interval $0<z<0.8$. In order to obtain that result
they applied a passive luminosity correction of $\Delta \log(M/L_B)=0.46 \pm
0.04$ to a sample drawn from COMBO--17 and DEEP2 \citep{davis03, faber07}. This
luminosity correction was obtained from the works by \citet{vdstanford} and
\citet{treu} and is lower than the one found a few years later by \citet{vd}
with a larger sample. This apparently small difference in $\Delta \log(M/L_B)$
has strong implications when studying the number density evolution of massive
red galaxies at the level of $\sim 1.5-2$ between $z=1$ and $z=0$. Also, the
number of galaxies in their COMBO--17 or DEEP2 samples is relatively small when
compared to the $\sim 7$ sq.~deg.~of the NOAO Deep Wide--Field Survey
\citep[NDWFS, ][]{jannuzi} used by \citet{brown}. \citet{cool} used a sample of luminous red galaxies (LRG)
from SDSS and spectroscopic follow--up from MMT to address a similar
question. They found no evolution in the number density of LRGs since
$z=0.9$. We believe that the main driver of the difference between Cool et
al.'s result and ours is, again, the correction for $M/L$ ratio
evolution. They used stellar population models from \citet{bc03} and
\citet{maraston}, which gives slower evolution than FP
zeropoint evolution constraints. 

In a recent work, \citet{mat} studied the number density evolution of massive
galaxies since $z\sim 1$ from a
sample drawn from UKIRT Infrared Digital Sky Survey and SDSS. Instead of
applying an average $M/L$ correction, they calculated
stellar masses for individual galaxies at all redshifts, finding a factor $\sim
4$ evolution at $M_*>10^{11}M_\sun$ (when adding-up their two mass
bins). Unlike our present work, they did not differenciate between red and
blue galaxies, but given the predominance of red galaxies at such high stellar
masses (they report a blue fraction $<30\%$ at all redshifts in the same mass
range) we believe that our factor $\sim 4$ evolution agrees remarkably well
with their result.

We now address the question of whether galaxy mergers between massive galaxies
can drive the observed number density evolution of massive red galaxies. For that purpose, we use the
result for the merger rate found in Section~\ref{sec:rate} and work under the
assumption that mergers between massive galaxies will produce
remnants with red optical colors. Dry (gas--free) mergers are likely to play an important role in the
build-up of the massive end of the red sequence \citep{vd_dry, bell06b}, and the
descendant systems of such interactions will clearly be red systems as
well. Moreover, mergers involving massive blue, disk--like galaxies are also
expected to produce red ellipticals as remnants. Major interactions,
independently of the color of the progenitors, are
expected to perturb the orbits of the stars in the galaxies and form
elliptical systems, but the observed fraction of spheroidal galaxies with blue optical
colors is a negligible fraction of the total at masses above $10^{11}M_\sun$
\citep[see][]{schawinski, kan, huertas}. It is likely that even those few systems will passively
evolve to the red sequence in $\lesssim 1$~Gyr, so, our assumption that merger
remnants with masses above such stellar mass will become red sequence galaxies
is justified.

We use the number density of massive systems at $z\sim 0.9$ that we have
obtained as the starting point and calculate the growth in the number of
massive red galaxies implied by our measurement of the merger rate. We show
the result of this exercise as the solid line in Fig.~\ref{fig:evol}.
 We stress that
the observed evolution (filled circles) and the one implied by our measurement (solid line) are {\it completely independent} except for the fact that we
use the observed density at $z\sim0.9$ to anchor the
evolution predicted by our close pair fractions. We find that mergers of
massive galaxies {\it can explain} the evolution in the observed number
density of massive red galaxies since $z=1$. In different words, the
  majority of galaxies observed
  today at the massive end of the red sequence have been
  assembled in the last 8~Gyrs through mergers of massive galaxies.

 We have used $\tau=0.5$~Gyrs, but using the $\tau \sim 1$~Gyr timescale from 
\citet{kw} produces a somewhat slower evolution that is still 
compatible within the error bars.

There are two caveats we would like to mention. Firstly, given the nature of our
method, some of the progenitor galaxies have masses above $10^{11}M_\sun$, so
strictly speaking they are not {\it newly formed} massive galaxies.  Second, 
because we adopt a lower mass limit of $5\times 10^{10}M_\sun$, we 
underestimate the number of major mergers that could lead to the formation
of a massive galaxy. For example, a pair with individual
masses $M_*=6\times 10^{10}M_\sun$ and
$M_*=4\times 10^{10}M_\sun$ would not make it into our pair
sample but would produce a major merger remnant of $10^{11} M_{\sun}$. 
Assuming that the merging population has a composition similar
to our overall sample, we estimate that this latter lower mass
limit issue has an impact  a factor $\sim 2$ larger on the creation of
$>10^{11}M_\sun$ galaxies than the overestimate caused by double counting
already massive
red galaxies (i.e., for every 10 massive galaxies we are
  counting as produced by mergers, there would be 1-2 which were already above
$10^{11}M_\sun$ and $\sim 3$ that we are not accounting for because one of the
progenitors has a mass below our mass limit of $5\times 10^{10}M_\sun$). This
means that our results are uncertain at the level of $\lesssim 20\%$ when counting
newly formed massive red galaxies. We do {\it not} apply this
correction to our results but would like to stress that, if anything, we
slightly underestimate the real creation rate of massive galaxies induced by merging.
 
 \citet{hopkins08} found by using HOD models a merger-driven evolution of a factor 3--4
  in the integrated stellar mass density in red elliptical galaxies
  between $z=1$ and $z=0$. Given that we focus on the massive end of the
  distribution, a quantitative comparison is impossible, but we argue that our
  results are in qualitative good agreement with the trend seen in that work
  and in \citet{hopkins06}, where they found some evolution in
  $M_*>10^{11}M_\sun$ red galaxies in the same redshift range by evolving the
  mass function of red galaxies in a manner compatible with the predicted
  merger rates.

\citet{white} studied the clustering of luminous red galaxies in
  the NDWFS in conjunction with HOD models, finding that $\sim 1/3$ of red
  luminous satellites present at $z=0.9$ have dissapeared by $z=0.5$. The most
  likely explanation to this phenomenon is that merging has taken place,
  decreasing the number of such galaxies, in good qualitative
  agreement with our result.

Also, in a recent work, \citet{arjen} studied the shape (axis ratio) distribution
of a large sample of quiescent galaxies drawn from the SDSS. Quiescent
galaxies with prominent disks are exceedingly rare at stellar masses above
$10^{11} M_{\sun}$, providing a strong indication that most, if not all,
massive quiescent galaxies form through major mergers. Our results are
consistent with this entirely indepent result, and, moreover, suggest that
most of this merger activity occurred over the past $\sim 8$~Gyrs, since $z=1$.

\section{Conclusions}

We have studied the impact of galaxy mergers on the evolution of massive
galaxies by using 2--point correlation functions to measure the fraction of galaxies in close pairs
from a sample of $\sim 18000$ galaxies more massive than
$5\times10^{10}M_\sun$ drawn from the COSMOS and COMBO--17 surveys and a pair
fraction estimate from SDSS. We have
also used constraints from the observed evolution of the fundamental plane
zeropoint to 
calculate the number density evolution of $M_* >10^{11}M_\sun$ red galaxies from
\citet{brown} LFs. Our main findings are:\\

%\begin{itemize}

1-.  The fraction of galaxies more massive than $5\times 10^{10}M_\sun$
in close pairs evolves as $F(z)=(0.0130\pm0.0019)\times (1+z)^{1.21\pm0.25}$
in the redshift range $0<z<1.2$. Assuming a merging timescale of
$\tau=0.5$~Gyr this implies that galaxies more massive than $10^{11}M_\sun$ have
undergone, on average, 0.5 (0.7) major mergers since $z=$~0.6 (1.2) involving
progenitors more massive than $5\times 10^{10}M_\sun$.\\

2-. The number density of $M_*>10^{11}M_\sun$ galaxies on the red sequence
  increases by a factor of $\sim 4$ between $z=1$ and $z=0$. This result
  depends strongly on the assumed evolution of the M/L ratio, which we have
  assumed here to be $\Delta \log(M/L_B)=0.555 \pm 0.042$ between $z=1$ and
  $z=0$ following recent estimates by \citet{vd}.\\

3-. The evolution implied by our measured merger rate is sufficient to explain this
  observed number density evolution of massive red galaxies
  since $z=1$; major mergers between massive ($M_*>5\times 10^{10}M_\sun$)
  galaxies are the main driver of the red sequence growth at its
  massive end.
%\end{itemize}

\acknowledgements

 A.\ R.\ R.\ and E.\ F.\ B.\ acknowledge support through 
the DFG Emmy Noether Programme. A.\ R.\ R.\ and R.\ E.\ S.\
acknowledge the Heidelberg--International Max Planck Research School
program. CW was supported by an STFC Advanced  
Fellowship.

\end{document}